\documentclass [12pt] {article}
\usepackage{a4}
 
\usepackage{amsfonts}
\usepackage{amsmath} 
\usepackage{graphics}
\usepackage{epsfig}

\author{T.~Leonhardt, R.~Manvelyan and W.~R\"uhl}
\title{The group approach to AdS space propagators: A fast algorithm}

\begin{document}

\newcommand {\eps}{\varepsilon}
\newcommand {\ti}{\tilde}
\newcommand {\D}{\Delta}
\newcommand {\G}{\Gamma}
\newcommand {\de}{\delta}
\newcommand {\al}{\alpha}
\newcommand {\la}{\lambda}
\newcommand {\si}{\sigma}
\newcommand {\vn}{\vec \nabla}
\newcommand {\vp}{\vec \partial}
\newcommand {\La}{\Lambda}
\newcommand {\ind}{\textrm{d}}

\thispagestyle{empty}

\noindent hep-th/0310063   \hfill  January 2004 \\                   
 
\noindent
\vskip3.3cm
\begin{center}
 
{\Large\bf  The group approach to AdS space propagators: A fast algorithm}
\bigskip\bigskip\bigskip
 
{\large Thorsten Leonhardt, Werner R\"uhl}
\medskip
 
{\small\it Fachbereich Physik, TU Kaiserslautern \\
       Postfach 3049, D-67653 Kaiserslautern, Deutschland}\\
{\small\tt tleon,ruehl@physik.uni-kl.de} \\
\bigskip
{\large Ruben Manvelyan}\\
\medskip
{\small\it Theoretical Physics Department, Yerevan Physics Institute \\
       Alikhanian Br. Str. 2, 375036 Yerevan, Armenia}\\
{\small\tt manvel@moon.yerphi.am } \\
\end{center}

\bigskip 
\begin{center}
{\sc Abstract}
\end{center}
\noindent
In this letter we show how the method of \cite{Leonhardt:2003qu} for the calculation of two-point functions in $d+1$-dimensional AdS space can be simplified. This results in an algorithm for the evaluation of the two-point functions as linear combinations of Legendre functions of the second kind. This algorithm can be easily implemented on a computer. For the sake of illustration, we displayed the results for the case of symmetric traceless tensor fields with rank up to $l=4$. 

\newpage


\section{Introduction}

In general (perturbative) quantum field theory, two-point functions arise in zeroth order as integral kernels inverting differential operators connected with the symmetries of the underlying spacetime. If interactions are included, the two-point functions receive corrections due to renormalization and we only have the K\"all\'en-Lehmann spectral representation with a positive spectral density function. Thus the form of the two-point function sensitively depends on the dynamics of the theory under consideration. In sharp contrast to this fact, in conformal field theory the form of the two-point function of a symmetric traceless tensor field of rank $l$ \footnote{Other fields can be treated similarly.} is fixed uniquely by the quantum numbers $\chi=[l, \D]$, where $\D$ denotes the conformal dimension of the corresponding tensor field. In general the value of $\D$ is changed by renormalization, but otherwise the two-point function is determined by the representation of the conformal group $G=SO(d+1,1)$ or $G=SO(d,2)$ for the case of Euclidean resp. Minkowskian signature. 

The constructive principle of the AdS/CFT correspondence (see \cite{Aharony:1999ti,D'Hoker:2002aw} and references therein), namely the possibility of  calculating correlation functions in $d$-dimensional conformal field theory from a field theory in $d+1$-dimensional Anti-de-Sitter space, can be turned around \cite{Dobrev:1998md}, at least for generic $\D$. Thus it is possible to calculate certain objects in the AdS field theory space from the corresponding objects in the CFT. In \cite{Leonhardt:2003qu} it is shown that this program can be carried out for two-point functions. Although this result is satisfactory from a conceptual point of view, the calculations turn out to be cumbersome, at least for the case of $l\ge2$. In this work, we shall circumvent these difficulties and show how the calculation can be simplified in an essential way. Finally we get an algorithm, which can be easily implemented on a computer \cite{homepage}.

With this algorithm at hand, we compute the bulk-to-bulk propagators for symmetric tensor fields with spin $0 \le l \le 4$. The cases $l = 0,1,2$ are well known in the literature and agreement is found. The explicit evaluation for tensor ranks $l=3$ and $4$ has not been performed up to now, and we do this as a simple application.

In principle, the same method can also be applied to higher $n$-point functions, i.e. take a conformal $n$-point function and convolute every argument with the appropriate bulk-to-boundary propagator. Especially the conformal three-point function of two scalar fields and one symmetric tensor field seems to be a promising candidate, due to its uniqueness. However, up to now we did not manage to solve the appearing convolution integral in an explicit way. Nevertheless, this is a representation of a three point function in $d+1$-dimensional AdS space in terms of a triple $d$-dimensional integral. 

Let us finally mention that the general method is in principle applicable to other spacetimes as well. The ingredients of our approach are the exactly known two-point function of one representation of the symmetry group $G$ and the knowledge of the intertwiners with an (at least partially) equivalent representation, but with a different representation space. If one is lucky enough, as we have been in the case of the AdS/CFT representation equivalence, one can find a pair of representation spaces $C_1,C_2$ of two equivalent irreps of the corresponding symmetry group, together with an intertwiner between $C_1$ and $C_2$ and a two-point function on one of the spaces, say $C_1$. By the procedure sketched below, the two-point function on $C_2$ can be constructed from the one on $C_1$.


\section{The method and its refinement}

In this work we will use Poincar\'e coordinates 
\begin{align}
x = (x_0, \vec x) = (x_0,x_1,\ldots,x_d) \in \mathbb{R}_+ \times \mathbb{R}^d
\end{align} 
for the points on $d+1$-dimensional AdS space. The conformal field theory lives on the $d$-dimensional boundary $\mathbb{R}^d$, which is located at $x_0=0$, therefore we denote the points on the boundary by $\vec x \in \mathbb{R}^d$. As usual, we understand $x_{12} = x_1 - x_2$, and extend this to the points on the boundary. Moreover, we write $\langle \cdot , \cdot \rangle$ and $(\cdot, \cdot)$ for the $d+1$- and $d$-dimensional standard scalar products and their extensions to the corresponding tensor spaces, respectively. By a theorem of Allen and Jacobson \cite{Allen:wd}, the propagators can be written as a polynomial in AdS invariant bitensors, which are related to simple geometric objects of the maximally symmetric space. The coefficients of these polynomials are functions of an AdS invariant distance, which we choose to be 
\begin{align}
\zeta := \frac{x_{10}^2 + x_{30}^2 + (\vec x_1 -\vec x_3)^2}{2 x_{10} x_{30}}.
\end{align} 

Let us briefly recall the computation method of the bulk-to-bulk propagator of a symmetric traceless tensor field of rank $l$ \cite{Leonhardt:2003qu}:
Consider the bulk-to-boundary propagator corresponding to a conformal rank $l$ symmetric traceless tensor of conformal dimension $\D$:
\begin{align}
K_\D^{(l)}(a, \vec b; x_1, \vec x_2) = \frac{x_{10}^{\D-l}}{(x_{12}^2)^\D} \biggl\{ \langle a, r(x_{12}) \vec b \rangle^l - \textrm{traces} \biggr\},
\end{align}
where we write $\la:= d-\D$ for the dimension of the shadow field, introduce the $(d+1)\times d$ matrix $r$ by
\begin{align}\label{r-mat}
r(x)_{\al, i} := \frac{2 x_\al x_i}{x^2} - \de_{\al, i},
\end{align}
and contract the indices $\al=0,\ldots,d; i=1,\dots,d$ with $a\in \mathbb{R}^{d+1}$ and $b\in \mathbb{R}^d$, respectively. In \cite{Leonhardt:2003qu} it is shown, that the convolution of $K_\D^{(l)}$ with $K_\la^{(l)}$ along the $d$ dimensional boundary leads to the sum of two bulk-to-bulk propagators of the respective parameters $\D$ and $\la$. These parameters arise from masses of the fields on the AdS space, e.g. in the scalar case we have
\begin{align}
m^2 = \D(\D-d).
\end{align}

The convolution in the earlier work was treated technically by writing $K_\D^{(l)}$ and $K_\la^{(l)}$ as a linear combination of differential operators $\mathcal{D}$ with respect to the exterior variables $x_1, x_3$ acting on $(x_{12}^2)^{-\la+p} (x_{32}^2)^{-\D+q}$, where $p,q \in \mathbb{Z}$. The differential operators $\mathcal{D}$ can then be pulled out of the integrals, which result in a linear combination of gaussian hypergeometric functions. These in turn can be  written in terms of Legendre functions of the second kind. Finally the differential operators $\mathcal{D}$ are evaluated, and the indices are grouped into polynomials of the bitensorial invariants $I_1,...I_4$ of the maximally symmetric space AdS. The traces are subtracted afterwards.

In this work, we introduce the following simplifications:
\begin{itemize}
\item The subtraction of traces is incorporated from the beginning by writing the Gegenbauer polynomial $C_l^{\mu-1}$ of order $l$ instead of the $l^{\textrm{th}}$ power of $\langle a, r(x) \vec b \rangle$ as well. Here we introduced the abbreviation $\mu:=d/2$ for convenience.
\item The elementary, but bothering rewriting of $K_\D^{(l)}$ in terms of differential operators $\mathcal{D}$ acting on powers of the distances $x_{12}^2, x_{23}^2$ is circumvented by a direct integration of the angular part of the $\vec x_2$ integral. The integrand of the remaining radial integral turns out to have a nice homogeneity property, which enables us to perform the radial integration. 
\item Moreover, we immediately obtain representations of the bulk-to-bulk propagators (including the shadow term) in terms of Legendre functions of the second kind, which can be shown to agree with the known bulk-to-bulk propagators using some identities for the Legendre functions. 
\end{itemize}

The main advantage of the new method lies in its algorithmic character. As we will see, the propagators can be written as a sum over $l+1$ terms. Each term factorizes into an AdS covariant bitensor and a linear combination of a special type of Legendre functions. The bitensor factors are polynomials of new basic bitensors $L_1, L_2, L_3, L_4$ which depend linearly on the bitensors $I_1, I_2, I_3, I_4$ of \cite{Leonhardt:2003qu}. The coefficients $R_{r_1,r_2,r_3,r_4}^{(l,k)} (\mu)$ in these polynomials are rational functions depending  only on $d=2 \mu$. We wrote a little Maple computer program (see \cite{homepage}), which computes the coefficients $R$, for which we did not find general analytic expressions. Finally we perform the AdS boundary limit of the propagators for arbitrary $l$ and express their normalization by comparison with the flat space CFT propagators. Thus we obtained a simple way of calculating two-point functions for symmetric traceless tensors, without the knowledge of their corresponding differential equations. We are aware of the fact that one expects the two-point functions to be expressible in terms of Legendre functions, due to the symmetry of the underlying space, and that the two-point functions are expected to be solutions of differential equations related to the Casimir operators of the symmetry group of the AdS space. However, in our approach we obtain explicit results for these propagators.


\section{The integration}
The propagators are integrals of the following form:
\begin{align}\label{propint}
A_{\D,\la}^{(l)} & := \int \langle a^l,K_\la^{(l)}(x_1, \vec x_2) \rangle_{i_1 \ldots i_l} \langle c^l, K_\D^{(l)}(x_3, \vec x_2) \rangle_{i_1 \ldots i_l} \ind^d x_2 \nonumber \\
& \, = n_l(\mu) \int \frac{x_{10}^{\la-l}}{(x_{12}^2)^\la} \frac{x_{30}^{\D-l}}{(x_{23}^2)^\D} \bigl(|A||C|\bigr)^l C_l^{\mu-1}(\frac{(\vec A, \vec C)}{|A||C|}) \ind^dx_2,
\end{align}
where 
\begin{align}
A_i &:= \langle a, r(x_{12})\rangle_i = 2 \langle a, x_{12} \rangle \frac{x_{12,i}}{x_{12}^2} -a_i, \nonumber \\
C_i &:= \langle c, r(x_{32})\rangle_i = 2 \langle c, x_{32} \rangle \frac{x_{32,i}}{x_{32}^2} -c_i,
\end{align}
and
\begin{align}
n_l (\mu) := \frac{2^{-l} l!}{(\mu-1)_l}.
\end{align}
Note that the Gegenbauer polynomial  $C_l^{\mu-1}$ corresponds to tracelessness in $d$ dimensions. Thanks to the property 
\begin{align}
^t r(x) r(x) = \textrm{id}_d
\end{align}
of the $r$-matrices (\ref{r-mat}), tracelessness with respect to the $d+1$ dimensional indices follows.

Now we exploit the symmetries of the problem: First we perform a rotation $R \in SO(d+1,1)$ of the coordinates, such that \footnote{Such a transformation exists, as can be easily seen by considering Euclidean AdS space as the set $\{ X \cdot X =1\} \in R^{d+1,1}$, where $\cdot$ denotes the Minkowskian scalar product (mostly plus) in this space.}
\begin{align}\label{coords1}
x_i := R(x_1)_i = R(x_3)_i,\quad i=1,\ldots,d.
\end{align}
Then we perform a translation in the $d$ dimensional integration variable $\vec x_2$ about $\vec x$, which is altogether the same as setting $\vec x_1=\vec x_3=0$ in (\ref{propint}). For convenience, we define 
\begin{align}\label{coords2}
\xi & := x_{12}= (x_{10},-\vec x_2), \nonumber \\
\eta & := x_{32} = (x_{30}, -\vec x_2).
\end{align}
Next we expand $A$ and $C$ in terms of $\vec x_2$:
\begin{align}
\vec A^2 &= \al_2 ( \vec a, \vec x_2)^2 + \al_1 ( \vec a, \vec x_2) + \al_0  \nonumber \\
\vec C^2 &= \beta_2 (\vec c, \vec x_2)^2 + \beta_1 (\vec c, \vec x_2) + \beta_0 \nonumber \\
(\vec A, \vec C) &= \gamma_{11} (\vec a, \vec x_2) (\vec c, \vec x_2) + \gamma_{10} (\vec a, \vec x_2) + \gamma_{01} (\vec c, \vec x_2) + \gamma_{00},
\end{align} 
where $(r^2 := \vec x_{2}^2)$
\begin{align}
\al_0 & = 4 (a_0 x_{10})^2 \frac{r^2}{\xi^4} + \vec a^2 \nonumber \\
\al_1 & = 4 \frac{a_0 x_{10}}{\xi^2} \Bigl( 1- 2\frac{r^2}{\xi^2} \Bigr) \nonumber \\
\al_2 & = \frac{4}{\xi^2} \Bigl(-1 + \frac{r^2}{\xi^2} \Bigr),
\end{align}
the coefficients $\beta_n$ are formed analogously, and for the $\gamma_n$ we have
\begin{align}
\gamma_{00} & = 4 a_0 c_0 x_{10} x_{30} \frac{r^2}{\xi^2 \eta^2} + (\vec a, \vec c) \nonumber \\
\gamma_{01} & = 2 \frac{a_0 x_{10}}{\xi^2} \Bigl( 1- 2\frac{r^2}{\eta^2} \Bigr) \nonumber \\
\gamma_{10} & = 2 \frac{c_0 x_{30}}{\eta^2} \Bigl( 1- 2\frac{r^2}{\xi^2} \Bigr) \nonumber \\
\gamma_{11} & = -\frac{2}{\xi^2 + \eta^2} (x_{10}^2 x_{30}^2 ).
\end{align}

We define the coefficients $\sigma_{nm}$ by 
\begin{align}
\sum_{k=0}^{[l/2]} \frac{(-\frac{l}{2})_k \,(\frac{1-l}{2})_k}{k! \,(2-\mu-l)_k} (\vec A, \vec C)^{l-2k} (\vec A^2 \vec C^2)^k = \sum_{n,m} \sigma_{nm} (\vec a, \vec x_2)^n (\vec c, \vec x_2)^m, 
\end{align}
and obtain for the integral (\ref{propint}) after all these manipulations 
\begin{align}
A_{\D,\la}^{(l)} & = x_{10}^{\la-l} x_{30}^{\D-l} n_l(\mu) \sum_{n,m} \int \ind^d x_2 \, \sigma_{nm} (\xi^2)^{-\la} (\eta^2)^{-\D}  (\vec a, \vec x_2)^n (\vec c, \vec x_2)^m.
\end{align}
We introduce spherical coordinates $(r, \Omega)$ for $\vec x_2$ and begin with the integration over the angles. The integral 
\begin{align}\label{sinangint}
r^{2 \nu} f_{nm} := \int \ind \Omega (\vec a, \vec x_2)^n (\vec c, \vec x_2)^m 
\end{align}
is one term in the binomial expansion of 
\begin{align}\label{sumint}
\int \ind \Omega \Bigl( (\vec a, \vec x_2) + (\vec c, \vec x_2) \Bigr)^{n+m} = |\vec a+ \vec c|^{2 \nu } r^{2 \nu} \frac{(\frac{1}{2})_\nu}{(\mu)_\nu} J_0(\mu),
\end{align}
where $\nu= \frac{n+m}{2}$ and $J_0(\mu) = 2 \pi^{\frac{d-1}{2}} \G(\frac{1}{2})/\G(\mu)$. Observe that $\nu$ is an integer, since only the terms with $n+m \in 2 \mathbb{Z}$ survive the angle integration. We expand the factor $|\vec a + \vec c|^\nu$ on the RHS of (\ref{sumint}) 
\begin{align}
\Bigl( \vec a^2 + 2 (\vec a, \vec c) + \vec c^2 \Bigr)^\nu = \sum_{k_1,k_2,k_3} \binom{\nu}{ k_1 k_2 k_3} (\vec a^2)^{k_1} (\vec c^2)^{k_2} \bigl(2 (\vec a, \vec c) \bigr)^{k_3},
\end{align}
and pick out the summand with the sought-for degree of homogeneity in $\vec a$ and $ \vec c$, i.e. 
\begin{align}
n= 2 k_1 + k_2, \quad m= 2 k_2 +k_3.
\end{align}
Then we obtain for (\ref{sinangint})
\begin{align}\label{effemen}
r^{2 \nu} f_{nm} & = J_0(\mu) \binom{2 \nu}{n}^{-1} r^{2\nu} \frac{(1/2)_\nu}{(\mu)_\nu} \sum_k \binom{\nu}{\frac{n-k}{2}, \frac{m-k}{2}, k} (\vec a^2)^{\frac{n-k}{2}} (\vec c^2)^{\frac{m-k}{2}} \bigl( 2 (\vec a, \vec c) \bigr)^k,
 \nonumber \\
& = J_0(\mu) \binom{2 \nu}{n}^{-1} r^{2\nu} \frac{(1/2)_\nu}{(\mu)_\nu} \sum_{k,p,q} \binom{\nu}{\frac{n-k}{2}-p, \frac{m-k}{2}-q, k, p, q} \nonumber \\
& \qquad \qquad \qquad \qquad \qquad \qquad ( - a_0^2)^{\frac{n-k}{2}-p} ( - c_0^2)^{\frac{m-k}{2}-q}  \bigl( 2 (\vec a, \vec c) \bigr)^k (a^2)^p (c^2)^q
\end{align}
where $k \in \mathbb{N}_0$ is such that $\frac{n-k}{2}, \frac{m-k}{2}
\in \mathbb{N}_0$ and we installed $a^2, c^2$ after the second sign of equality by
\begin{align}
(\vec a^2)^{\frac{n-k}{2}} = ( a^2 -a_0^2)^{\frac{n-k}{2}} = \sum_p \binom{\frac{n-k}{2}}{p} (a^2)^p (-a_0^2)^{\frac{n-k}{2}-p}
\end{align}
(similarly for $c^2$). In the special coordinate system (\ref{coords1},\ref{coords2}) the variables in the last line of (\ref{effemen}) are expressed (see below) in terms of the bitensors 
\begin{align}
L_1 & := \frac{(\vec a, \vec c)}{x_{10} x_{30}} = - I_1 - \frac{\zeta}{1-\zeta^2} I_2 \Bigr \rvert_{\vec x_1 = \vec x_3 = 0} \nonumber \\
L_2 & := \frac{a_0 c_0}{x_{10} x_{30}} = \frac{1}{1-\zeta^2} I_2 \Bigr \rvert_{\vec x_1 = \vec x_3 = 0} \nonumber \\
L_3 & := \frac{a^2 c_0^2 + c^2 a_0^2}{x_{10}^2 x_{30}^2} = -\frac{1}{1-\zeta^2} I_3 \Bigr \rvert_{\vec x_1 = \vec x_3 = 0} \nonumber \\
L_4 &:= \frac{a^2 c^2}{x_{10}^2 x_{30}^2} = I_4,
\end{align}
where $L_{3,4}$ are reserved for the trace terms and the $I_j$ are the covariant bitensors of \cite{Leonhardt:2003qu}.

After the angle integration we can now come to the integration over
the radial variable:
\begin{align}\label{eqnforA}
A_{\D,\la}^{(l)} = x_{10}^{\la-l} x_{30}^{\D-l} n_l(\mu) \sum_{n,m}  f_{nm} \int
\ind r\, r^{d-1+2\nu} \sigma_{nm} \bigl( \xi^2 \bigr)^{-\la} \bigl(
\eta^2 \bigr)^{-\D} 
\end{align}
We have found that the sum over $n,m$ can be written in
the following way (we checked this up to $l=8$):
\begin{align}
\sum_{n,m}  f_{nm} \sigma_{nm} r^{2\nu} = \sum_{s=0}^{l}  B_s^{(l)}
\Bigl( \frac{r^2}{\xi^2 \eta^2} \Bigr)^s,
\end{align}
where the coefficients $B_s^{(l)}$ depend solely on $x_{10}$ and
$x_{30}$. Thus we have to do the integral
\begin{align}
K_s &= \int_0^\infty \ind r \,r^{d+2s-1} \bigl( \xi^2 \bigr)^{-\la-s}
\bigl( \eta^2 \bigr)^{-\D -s} \nonumber \\
&= \frac{1}{2} \int_0^\infty \ind t \, t^{\mu+s-1} \bigl(1+t
\bigr)^{-\D-s} \bigl( 1+ \rho t \bigr)^{-\la-s} \bigl( x_{10}^2
\bigr)^{-\la-s} \bigl( x_{30}^2 \bigr)^{-\D+\mu},
\end{align}
where the substitution 
\begin{align}
r^2 &= x_{30}^2 t \nonumber \\
\rho &= \frac{x_{30}^2}{x_{10}^2}
\end{align}
was used. This integral can be found in (3.197,5 of \cite{GR}), giving
\begin{align}
K_s & = \frac{1}{2} \bigl( x_{10}^2 \bigr)^{-\la-s} \bigl( x_{30}^2
\bigr)^{-\D+\mu} B(\mu+s, \mu+s) \,\!_2F_1 \Bigl[{\la+s,\mu+s \atop 2\mu+2s}
; 1-\rho \Bigr] \nonumber \\
& = \frac{1}{2} \bigl( x_{10}^2 \bigr)^{-\mu-s} \frac{\G(\mu+s)
\G(\la-\mu)}{\G(\la+s)} \,\!_2F_1 \Bigl[ {\D+s, \mu+s \atop
\mu-\la+1} ; \rho \Bigr] + \{ \D \leftrightarrow \la \}.
\end{align}
The second summand, the ``shadow term'', can be discarded, and the
first term, the direct term, can be written in terms of Legendre functions of the second
kind. However, to get the connection with \cite{Leonhardt:2003qu}, we use a quadratic transformation for the hypergeometric function (eq. 9.134,2 of \cite{GR}) and represent $K_s$ in terms of 
\begin{align}
\ti \La_s ( \zeta ) & := \frac{(\mu)_s}{(\la)_s \G(\D+s)}
\La_{-s,-s}(\zeta) \nonumber \\
&\, = \frac{(\mu)_s}{(\la)_s} (2 \zeta)^{-(\D+s)}
\,\!_2F_1\Bigl[ {\frac{\D+s}{2}, \frac{\D+s+1}{2} \atop \D-\mu+1} \;
;\zeta^{-2} \Bigr],
\end{align}
where $\zeta$ in the coordinates (\ref{coords1},\ref{coords2}) is
\begin{align}
\zeta = \frac{x_{10}^2+x_{30}^2}{2 x_{10} x_{30}} = \frac{1+\rho}{2 \sqrt{\rho}}.
\end{align} 
Thus we finally arrive at 
\begin{align}
A_{\D,\la}^{(l)} = (x_{10} x_{30})^{-l} \frac{1}{2} J_0(\mu) n_l(\mu) \frac{\G(\mu)
\G(\la-\mu)} {\G(\la)} \sum_{s=0}^l B_s^{(l)} \ti
\La_s(\zeta) + \textrm{shadow term}.
\end{align}

We wrote a Maple computer program for the $B_s^{(l)}$, which are polynomials in $\zeta$ of degree $s$. We observed that the Legendre functions can be combined into $l+1$ expressions containing 
\begin{align}
Q_k(\zeta) = \sum_{s = 0}^{l-k} \frac{(-l+k)_s (k+1)_s}{(\mu+k)_s s!} (2 \zeta)^s \ti \La_{s+k} (\zeta),
\end{align}
where $k \in \{0,1,\ldots,l\}$, so that each $Q_k$ is multiplied with a bitensor. Namely, the direct term of $A_{\D,\la}^{(l)}$, eq. (\ref{eqnforA}), is 
\begin{align}\label{propall}
A_{\la,\D}^{(l)} \Bigr \rvert_{\textrm{direct}} = \kappa_l(\mu,\D) \sum_{k=0}^l Q_k(\zeta) \sum_{r_1,\ldots,r_4} R_{r_1,r_2,r_3,r_4}^{(l,k)}(\mu) L_1^{r_1} L_2^{r_2} L_3^{r_3} L_4^{r_4},
\end{align}
where $\kappa_0$ is a nonvanishing constant
\begin{align}\label{normkappa}
\kappa_l(\mu,\D) = \frac{1}{2} J_0(\mu) n_l(\mu) B(\mu,\mu-\D).
\end{align} 
For $0 \le l \le 4$ we list up the coefficients $R$ in the Appendix. 

Here we quote a few properties of the $R$. They are rational functions of $d= 2\mu$. Moreover,
\begin{align}
R_{l,0,0,0}^{(l,0)} =1
\end{align}
and $R$ is different from zero only if 
\begin{align}
r_1+r_2+2r_3+2r_4 = l \nonumber \\
l-k+r_1 \in 2 \mathbb{N}_0.
\end{align}
Then the sign of $R$ can be expressed as 
\begin{align}
\textrm{sign} R_{r_1, \ldots, r_4}^{(l,k)} = (-1)^{\frac{1}{2}(l-k-r_1)+r_3}.
\end{align}

Finally we want to investigate the AdS boundary limit in an arbitrary coordinate system. We set for simplicity 
\begin{align}
a_0=c_0=0
\end{align}
and compute the leading term of (\ref{propall}) for $x_{10} \to 0, x_{30} \to 0$. Then we expect the limit to be equal to the conformal two-point function of a symmetric traceless tensor field times a power of $x_{10} x_{30}$
\begin{align}\label{expbehav}
\lim_{x_{10},x_{30}\to 0} (x_{10} x_{30})^{-(\D-l)} A_{\la,\D}^{(l)} \Bigr \rvert_{\textrm{direct}} = \mathcal{N}_l (\mu,\D) (\vec x_{13}^2)^{-\D} \Bigl( \al(\vec a, \vec c)^l - \textrm{traces} \Bigr),
\end{align}
where
\begin{align}
\al(\vec a, \vec c) = -2 \frac{(\vec a, \vec x_{13}) (\vec c, \vec x_{13})}{ \vec x_{13}^2} + (\vec a,\vec c).
\end{align}

Among the basic bitensors only $L_1$ and $L_4$ contribute at $x_{10}=x_{30}=0$, with
\begin{align}
L_1 \sim \frac{\al(\vec a,\vec c)}{x_{10} x_{30}}, \quad L_4 \sim \frac{\vec a^2 \vec c^2}{x_{10}^2 x_{30}^2}
\end{align}
for $x_{10}, x_{30} \to 0$, which is the same as $\zeta \to \infty$. It is easy to see that 
\begin{align}
Q_k(\zeta) = O(\zeta^{-\D-k}) \qquad (\zeta \to \infty)
\end{align}
and especially
\begin{align}
Q_0(\zeta) \sim \frac{\la-1}{\la+l-1} (2 \zeta)^{-\D} \qquad (\zeta \to \infty).
\end{align}
Thus only $k=0$ survives and the expected behaviour (\ref{expbehav}) is reproduced with (including the constant $\kappa_0$, see (\ref{normkappa}))
\begin{align}
\mathcal{N}_l (\mu,\D) = \frac{1}{2} J_0(\mu) n_l(\mu) \frac{\la-1}{\la+l-1} B(\mu,\mu-\D)
\end{align}
and 
\begin{align}\label{alminustr}
\al(\vec a,\vec c)^l - \textrm{traces} = \sum_{k=0}^{l/2} R_{l-2k,0,0,k}^{(l,0)} \al(\vec a, \vec c)^{l-2k} (\vec a^2 \vec c^2)^k.
\end{align}
Since the LHS of (\ref{alminustr}) is given by the Gegenbauer polynomial $C_l^{\mu-1} (\frac{\al(\vec a,\vec c)}{|\vec a| |\vec c|})$, we get the same coefficients as in (\ref{propint}), i.e. 
\begin{align}
R_{l-2k,0,0,k}^{(l,0)} & = \frac{(-\frac{l}{2})_k (\frac{1-l}{2})_k}{k! (2-\mu-l)_k}= \frac{l!}{2^{2k} k! (l-2k)! (2-\mu-l)_k},
\end{align}
which indeed agrees with the cases considered in the Appendix.


\section{Appendix}

The coefficients $R_{r_1,r_2,r_3,r_4}^{(l,k)}$ are 
\begin{align}
l=0 :& \qquad R_{0,0,0,0}^{(0,0)} = 1 \\
l=1 :& \qquad R_{1,0,0,0}^{(1,0)} = 1 \\
 & \qquad R_{0,1,0,0}^{(1,1)} =4 
\end{align}
\begin{align}
l=2 :& \qquad R_{2,0,0,0}^{(2,0)} = 1 \\
 & \qquad R_{0,2,0,0}^{(2,0)} = -  R_{0,0,1,0}^{(2,0)}= R_{0,0,0,1}^{(2,0)} = - \frac{1}{2\mu}\\
 & \qquad R_{1,1,0,0}^{(2,1)} = \frac{4}{\mu^2} (\mu+1)(2\mu-1) \\
 & \qquad R_{2,0,0,0}^{(2,2)} = -\frac{4}{\mu^2 (\mu+1)} \\
 & \qquad R_{0,2,0,0}^{(2,2)} = \frac{2}{\mu^2 (\mu+1)} (8 \mu^3+12\mu^2+2\mu-3) \\
 & \qquad R_{0,0,1,0}^{(2,2)} = - \frac{2}{\mu^2 (\mu+1)} (4\mu^2+4\mu-1) \\
 & \qquad R_{0,0,0,1}^{(2,2)} = \frac{2}{\mu^2 (\mu+1)} (2\mu+1) 
\end{align}
\begin{align}
l=3 :& \qquad R_{3,0,0,0}^{(3,0)} = 1 \\
 & \qquad R_{1,2,0,0}^{(3,0)} = - R_{1,0,1,0}^{(3,0)} = R_{1,0,0,1}^{(3,0)} = -\frac{3}{2(\mu+1)} \\
 & \qquad R_{2,1,0,0}^{(3,1)} = \frac{12}{\mu+1}(\mu+2) \\
 & \qquad R_{0,3,0,0}^{(3,1)} = - R_{0,1,1,0}^{(3,1)} = R_{0,1,0,1}^{(3,1)} = - \frac{6}{(\mu)_2}(\mu+2) \\
 & \qquad R_{3,0,0,0}^{(3,2)} = -\frac{12}{\mu(\mu+1)^2} \\
 & \qquad R_{1,2,0,0}^{(3,2)} = \frac{6}{\mu(\mu+1)^2}(8\mu^3+28\mu^2+18\mu-15) \\
 & \qquad R_{1,0,1,0}^{(3,2)} = -\frac{6}{\mu(\mu+1)^2}(4\mu^2+8\mu-3) \\
 & \qquad R_{1,0,0,1}^{(3,2)} = \frac{6}{\mu(\mu+1)^2}(2\mu+1) \\
 & \qquad R_{2,1,0,0}^{(3,3)} = -\frac{144}{\mu(\mu+1)^2} \\
 & \qquad R_{0,3,0,0}^{(3,3)} = \frac{8}{\mu(\mu+1)^2}(8\mu^3+28\mu^2+26\mu-3) \\
 & \qquad R_{0,1,1,0}^{(3,3)} = - \frac{24}{\mu(\mu+1)^2}(4\mu^2+8\mu+1) \\
 & \qquad R_{0,1,0,1}^{(3,3)} = \frac{72}{\mu(\mu+1)^2}(2\mu+1) 
\end{align}
\begin{align}
l=4 :& \qquad R_{4,0,0,0}^{(4,0)} = 1 \\
 & \qquad R_{2,2,0,0}^{(4,0)} = - R_{2,0,1,0}^{(4,0)} = R_{2,0,0,1}^{(4,0)} = -\frac{3}{\mu+2} \\
 & \qquad R_{0,4,0,0}^{(4,0)} = R_{0,0,2,0}^{(4,0)} = R_{0,0,0,2}^{(4,0)} = \frac{3}{4(\mu+1)_2} \\
 & \qquad R_{0,2,1,0}^{(4,0)} = - R_{0,2,0,1}^{(4,0)} = R_{0,0,1,1}^{(4,0)} = - \frac{3}{2(\mu+1)_2} \\
 & \qquad R_{3,1,0,0}^{(4,1)} = 8\frac{(2\mu+1)(\mu+3)}{\mu(\mu+2)} \\
 & \qquad R_{1,3,0,0}^{(4,1)} = - R_{1,1,1,0}^{(4,1)} = R_{1,1,0,1}^{(4,1)} = - \frac{12}{(\mu)_3} (2\mu+1)(\mu+3) \\
 & \qquad R_{4,0,0,0}^{(4,2)} = -\frac{24}{(\mu)_3} \\
 & \qquad R_{2,2,0,0}^{(4,2)} = \frac{12}{(\mu)_3(\mu+1)} (8\mu^4 +52\mu^3+98\mu^2+39\mu+3) \\
 & \qquad R_{2,0,1,0}^{(4,2)} = - \frac{12}{(\mu)_3(\mu+1)} (4\mu^3+16\mu^2+9\mu+3) \\
 & \qquad R_{2,0,0,1}^{(4,2)} = \frac{12}{(\mu)_3(\mu+1)} (2\mu^2+3\mu+3) \\
 & \qquad R_{0,4,0,0}^{(4,2)} = - \frac{6}{(\mu)_3(\mu+1)} (8\mu^3+52\mu^2+98\mu+39) \\
 & \qquad R_{0,2,1,0}^{(4,2)} = \frac{12}{(\mu)_3(\mu+1)} (4\mu^3+28\mu^2+57\mu+24) \\
 & \qquad R_{0,2,0,1}^{(4,2)} = - \frac{12}{(\mu)_3(\mu+1)} (4\mu^3+26\mu^2+50\mu+21) \\
 & \qquad R_{0,0,2,0}^{(4,2)} = - \frac{6}{(\mu)_3(\mu+1)} (4\mu^2+16\mu+9) \\
 & \qquad R_{0,0,1,1}^{(4,2)} = \frac{12}{(\mu)_3(\mu+1)} (2\mu^2+9\mu+6) \\
 & \qquad R_{0,0,0,2}^{(4,2)} = - \frac{6}{(\mu)_3(\mu+1)} (2\mu+3) \\
 & \qquad R_{3,1,0,0}^{(4,3)} = - 288 \frac{(2\mu+1)(\mu+3)}{(\mu)_3(\mu+1)_2} \\
 & \qquad R_{1,3,0,0}^{(4,3)} = 16 \frac{(2\mu+1)(\mu+3)}{(\mu)_3(\mu+1)_2} (8 \mu^3+44\mu^2+58\mu-23) \\
 & \qquad R_{1,1,1,0}^{(4,3)} = -48 \frac{(2\mu+1)(\mu+3)}{(\mu)_3(\mu+1)_2} (4\mu^2+12\mu+1) \\
 & \qquad R_{1,1,0,1}^{(4,3)} = 144 \frac{(2\mu+1)(\mu+3)}{(\mu)_3(\mu+1)_2} (2\mu+1)
\end{align}
\begin{align}
 & \qquad R_{4,0,0,0}^{(4,4)} = 288 \frac{1}{(\mu)_4(\mu+1)_2} \\
 & \qquad R_{2,2,0,0}^{(4,4)} = - \frac{288}{(\mu)_4(\mu+1)_2} (8\mu^3+52\mu^2+98\mu+39) \\
 & \qquad R_{0,4,0,0}^{(4,4)} = \frac{4}{(\mu)_4(\mu+1)_2} (64 \mu^6+768\mu^5+3664\mu^4+8640\mu^3 \nonumber \\
& \qquad \qquad \qquad \qquad \qquad \qquad \qquad \qquad \qquad +10060\mu^2+4992\mu+927) \\
 & \qquad R_{2,0,1,0}^{(4,4)} = \frac{288}{(\mu)_4(\mu+1)_2} (4\mu^2+16\mu+9) \\
 & \qquad R_{2,0,0,1}^{(4,4)} = - \frac{288}{(\mu)_4(\mu+1)_2} (2\mu+3) \\
 & \qquad R_{0,2,1,0}^{(4,4)} = -\frac{24}{(\mu)_4(\mu+1)_2} (32\mu^5+304\mu^4+1072\mu^3+1688\mu^2+1122\mu+279) \\
 & \qquad R_{0,2,0,1}^{(4,4)} = \frac{24}{(\mu)_4(\mu+1)_2} (80\mu^4+544\mu^3+1240\mu^2+1088\mu+321) \\
 & \qquad R_{0,0,2,0}^{(4,4)} = \frac{12}{(\mu)_4(\mu+1)_2} (16\mu^4+128\mu^3+344\mu^2+352\mu+129) \\
 & \qquad R_{0,0,1,1}^{(4,4)} = - \frac{72}{(\mu)_4(\mu+1)_2} (2\mu+3)(4\mu^2+16\mu+11) \\
 & \qquad R_{0,0,0,2}^{(4,4)} = \frac{36}{(\mu)_4(\mu+1)_2} (2\mu+3)(2\mu+5)
\end{align}


\begin{thebibliography}{99}

\bibitem{Aharony:1999ti}
O.~Aharony, S.~S.~Gubser, J.~M.~Maldacena, H.~Ooguri and Y.~Oz,
``Large N field theories, string theory and gravity,''
Phys.\ Rept.\  {\bf 323} (2000) 183
[arXiv:hep-th/9905111].

\bibitem{D'Hoker:2002aw}
E.~D'Hoker and D.~Z.~Freedman,
``Supersymmetric gauge theories and the AdS/CFT correspondence,''
arXiv:hep-th/0201253.

\bibitem{Dobrev:1998md}
V.~K.~Dobrev,
``Intertwining operator realization of the AdS/CFT correspondence,''
Nucl.\ Phys.\ B {\bf 553} (1999) 559
[arXiv:hep-th/9812194].

\bibitem{Leonhardt:2003qu}
T.~Leonhardt, R.~Manvelyan and W.~R\"uhl,
``The group approach to AdS space propagators,''
Nucl.\ Phys.\ B {\bf 667} (2003) 413
[arXiv:hep-th/0305235].

\bibitem{Allen:wd}
B.~Allen and T.~Jacobson,
``Vector Two Point Functions In Maximally Symmetric Spaces,''
Commun.\ Math.\ Phys.\  {\bf 103} (1986) 669.


\bibitem{homepage}
See the program on our website ``http://theorie.physik.uni-kl.de/w\_ruehl/ prop81.html''.

\bibitem{GR}
I.~S.~Gradshteyn and I.~M.~Ryzhik,
``Table of Integrals, Series, and Products'', \textit{Academic Press
(1965)}. 



\end{thebibliography}
\end{document}